\documentclass[3p,times,procedia]{elsarticle}
\usepackage{nupha_ecrc}

\volume{00}

\firstpage{1}

\journalname{Nuclear Physics A}

\runauth{}

\jid{nupha}

\jnltitlelogo{Nuclear Physics A}

\usepackage{amssymb}

\usepackage[figuresright]{rotating}

\begin{document}

\begin{frontmatter}



\dochead{XXVIth International Conference on Ultrarelativistic Nucleus-Nucleus Collisions\\ (Quark Matter 2017)}

\title{A hybrid approach to relativistic heavy-ion collisions at the RHIC BES energies}


\author[a]{Chun Shen}
\author[b]{Gabriel Denicol}
\author[c]{Charles Gale}
\author[c]{Sangyong Jeon}
\author[d]{Akihiko Monnai}
\author[a]{Bjoern Schenke}
\address[a]{Physics Department, Brookhaven National Laboratory, Upton, NY 11973, USA}
\address[b]{Instituto de F\'{i}sica, Universidade Federal Fluminense, UFF, Niter\'{o}i, 24210-346, RJ, Brazil}
\address[c]{Department of Physics, McGill University, 3600 University Street, Montreal, QC, H3A 2T8, Canada}
\address[d]{Institut de physique theorique, Universite Paris Saclay, CNRS, CEA, F-91191 Gif-sur-Yvette, France}

\begin{abstract}

Using a hybrid (viscous hydrodynamics + hadronic cascade) framework, we model the bulk dynamical evolution of relativistic heavy-ion collisions at Relativistic Heavy Ion Collider (RHIC) Beam Energy Scan (BES) collision energies, including the effects from non-zero net baryon current and its dissipative diffusion. This framework is in full (3+1)D, which allows us to study the non-trivial longitudinal structure and dynamics of the collision systems, for example baryon stopping and transport, as well as longitudinal fluctuations. For the first time, the quantitative effect of net-baryon diffusion on hadronic observables is addressed. Finally, we propose a dynamical initialization scheme to study the importance of the pre-equilibrium stage at the RHIC BES energies.
\end{abstract}


\end{frontmatter}


\section{Introduction}
\label{intro}

The Beam Energy Scan (BES) program at RHIC provides us with a unique access to study the phase diagram of nuclear matter \cite{Adamczyk:2017iwn}. The wealth of hadronic measurements can help us to understand the transport properties of the quark-gluon plasma (QGP) at finite baryon density. Modelling the dynamical evolution of the collisions at the BES energies plays a critical role in this effort. Relativistic hydrodynamics is a successful phenomenological model for heavy-ion collisions at high collision energies \cite{Heinz:2013th,Gale:2013da}. In this work, we extend the hybrid framework (viscous hydrodynamics + hadronic cascade) to lower collision energies by including the propagation of net baryon current as well as its dissipative net baryon diffusion. This advanced framework opens a new opportunity to extract the heat conductivity of the QGP in a baryon rich environment.

One of the ultimate goals of the RHIC BES program is to find the location of the critical point. This requires knowledge of the trajectories of the fireballs in the phase diagram. Our hybrid framework can provide these trajectories and thus can act as a precise compass for searching the critical point in the QCD phase diagram. 

\section{The hybrid framework}
\label{model}

The initial entropy density and net baryon density profiles at $\tau_0$ are given by the Monte-Carlo Glauber model. In this work, we use event-averaged nuclear thickness functions in the transverse plane $T_{A, B} (x, y)$ folded with envelope functions along the longitudinal direction
\begin{equation}
s(x, y, \eta; \tau_0) = \frac{s_0}{\tau_0} \sum_{i=\{A, B\}} f^{s}_i (\eta) T_i (x, y), \qquad \rho_B(x, y, \eta; \tau_0) = \frac{1}{\tau_0} \sum_{i=\{A, B\}} f^{\rho_B}_i (\eta) T_i (x, y).
\label{eq1}
\end{equation}
The envelope function for entropy density $f^{s}_i (\eta)$ is chosen to be as in Ref. \cite{Bozek:2010vz}. The envelope function for net baryon density is assumed as a normalized asymmetric gaussian which peaks at forward or backward rapidity \cite{Shen:2017xxx}. These initial density profiles are evolved with viscous hydrodynamics
\begin{equation}
\partial_\mu T^{\mu\nu} = 0 \qquad \mbox{and} \qquad \partial_\mu J^{\mu} = 0,
\end{equation}
where the system's energy momentum tensor $T^{\mu\nu}$ and net baryon current $J^\mu$ are 
\begin{equation}
T^{\mu\nu} = eu^\mu u^\nu - P \Delta^{\mu\nu} + \pi^{\mu\nu}, \qquad J^{\mu} = \rho_B u^\mu + q^\mu.
\label{eq3}
\end{equation}
Here $\Delta^{\mu\nu} = g^{\mu\nu} - u^\mu u^\nu$ is the spatial projection operator. These hydrodynamic equations need to be solved together with a given equation of state (EoS) $P(e, \rho_B)$. In the QGP phase, the $\rho_B$-dependence of our EoS is constructed using Taylor expansion with the Lattice QCD susceptibility results up to $\mathcal{O}(\mu_B^4)$.  This is combined with a hadron resonance gas EoS in the hadronic phase. 
In Eq.~(\ref{eq3}) we include the net baryon diffusion current $q^\mu$, which allows the net baryon current to flow differently compared to the energy flow $u^\mu$. The shear stress tensor $\pi^{\mu\nu}$ and $q^\mu$ are evolved with second order Israel-Stewart type of equations
\begin{equation}
\Delta^{\mu\nu}_{\alpha \beta} D \pi^{\alpha \beta} = - \frac{1}{\tau_\pi} (\pi^{\mu\nu} - 2 \eta \sigma^{\mu\nu}) - \frac{\delta_{\pi\pi}}{\tau_{\pi}} \pi^{\mu\nu} \theta - \frac{\tau_{\pi\pi}}{\tau_\pi}\pi^{\lambda \langle} \sigma^{\nu \rangle}\,_\lambda + \frac{\phi_7}{\tau_\pi} \pi^{\langle \mu}\,_\alpha \pi^{\nu \rangle \alpha}
\label{eq4}
\end{equation}
\begin{equation}
\Delta^{\mu\nu} D q_\nu = - \frac{1}{\tau_q} \left(q^\mu - \kappa_B \nabla^\mu \frac{\mu_B}{T} \right) - \frac{\delta_{qq}}{\tau_q} q^\mu \theta - \frac{\lambda_{qq}}{\tau_q} q_\nu \sigma^{\mu\nu}.
\label{eq5}
\end{equation}
The double projection operator is defined as $\Delta^{\alpha \beta}_{\mu\nu} = \frac{1}{2}(\Delta^\alpha_\mu \Delta^\beta_\nu + \Delta^\alpha_\nu \Delta^\beta_\mu) - \frac{1}{3}\Delta^{\alpha \beta} \Delta_{\mu\nu}$. The shear viscosity is chosen as $\eta T/(e + P) = 0.08$ and $\tau_\pi = 0.4/T$. The net baryon diffusion constant $\kappa_B$ takes the form derived from the relaxation time approximation \cite{Jaiswal:2015mxa}
\begin{equation}
\kappa_B = \frac{C_B}{T} \rho_B \left(\frac{1}{3} \coth(\alpha_B) - \frac{\rho_B T}{e + P} \right) \qquad \mbox{and} \qquad \tau_q = \frac{C_B}{T}.
\label{eq6}
\end{equation}
The coefficient $C_B$ is a constant. In this work, we will vary the value of $C_B$ to study the effect of net baryon number diffusion on hadronic observables. The detailed expression for the other second order transport coefficients in Eqs.~(\ref{eq4}) and (\ref{eq5}) are listed in Ref.~\cite{Shen:2017xxx}. When the local energy density drops to $e_\mathrm{sw} = 0.3$ GeV/fm$^3$, individual fluid cells are converted into particle using the Cooper-Frye prescription
\begin{equation}
E \frac{d^3 N_i}{d^3 p}(x^\mu) = \frac{g_i}{(2\pi)^3}p^\mu \Delta^3 \sigma_\mu(x^\mu) \bigg(f^\mathrm{eq}_i(E, T, \mu_B) + \delta f^\mathrm{shear}_i(E, T, \mu_B, \pi^{\mu\nu}) + \delta f^\mathrm{diffusion}_i(E, T, \mu_B, q^\mu) \bigg) \bigg\vert_{E = p \cdot u}.
\label{eq.7}
\end{equation}
The out-of-equilibrium correction for net baryon diffusion is derived from the relaxation time approximation \cite{Jaiswal:2015mxa}
 \begin{equation}
\delta f^\mathrm{diffusion}_i (x, p) = f^\mathrm{eq}_i(x, p) (1 \pm f^\mathrm{eq}_i(x, p)) \left(\frac{\rho_B}{e + P} - \frac{b_i}{E} \right) \frac{p^{\mu} q_\mu}{\kappa_B/\tau_q}.
\end{equation}
The shear viscous correction is $\delta f^\mathrm{shear}_i(E, T, \mu_B, \pi^{\mu\nu}) = f^\mathrm{eq}_i(x, p) (1 \pm f^\mathrm{eq}_i(x, p))\frac{p_\mu p_\nu \pi^{\mu\nu}}{2 T^2 (e + P)}$. After the particle conversion, we feed them into a hadronic cascade model UrQMD \cite{Bass:1998ca} to further simulate the transport dynamics in the dilute hadronic phase.

\section{Results and discussions}
\label{results}

\begin{figure}[ht!]
  \centering
  \begin{tabular}{cc}
   \includegraphics[width=0.4\linewidth]{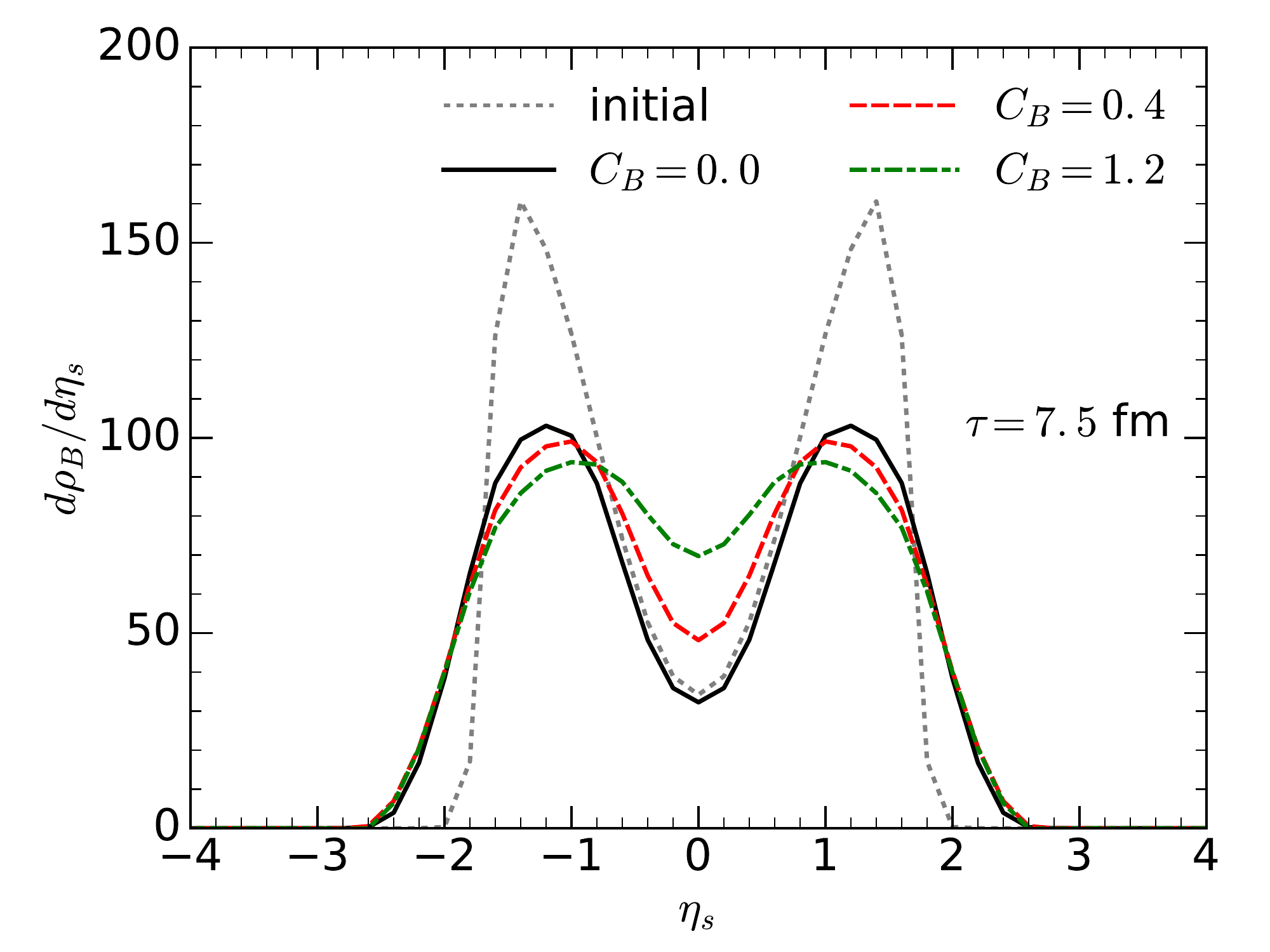} &
   \includegraphics[width=0.4\linewidth]{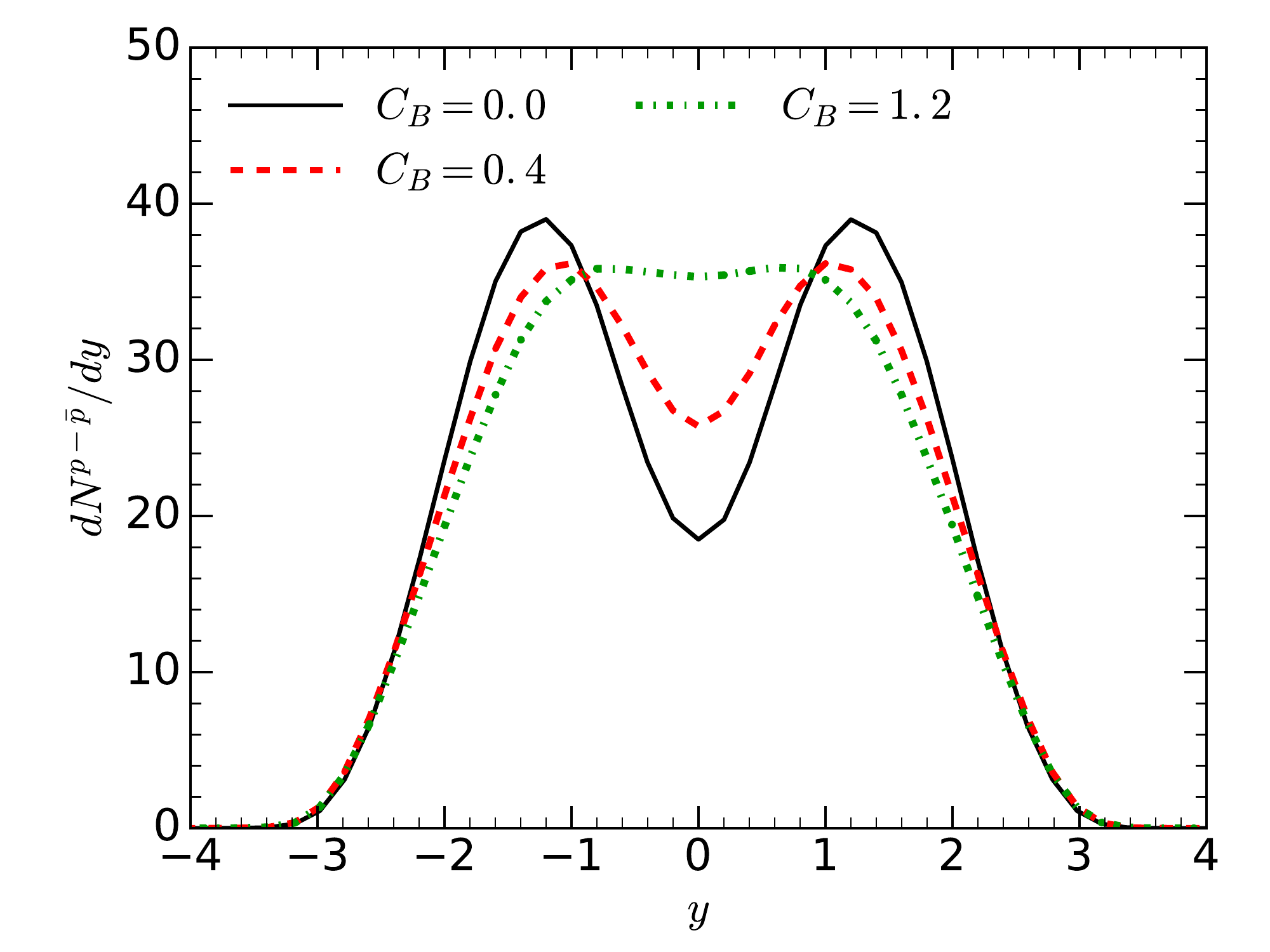}
   \end{tabular}
   \caption{{\it Left panel:} The space-time rapidity distribution of the net baryon density at a fixed proper time $\tau = 7.5$ fm/c during the hydrodynamic evolution with different net baryon diffusion constants.
{\it Right panel:} The effect of net baryon diffusion on the net proton rapidity distribution in 0-5\% Au+Au collisions at 19.6 GeV.}
  \label{fig1}
\end{figure}

Fig.~\ref{fig1} shows the baryon diffusion effect on the evolution of the net baryon density and how it affects the rapidity distribution for net protons. The left panel of Fig.~\ref{fig1} shows that the non-boost-invariant hydrodynamic evolution changes the longitudinal shape of the net baryon density distribution. The tails of the net baryon density profile get broadened and extend to more forward and backward space-time rapidity. This is because the pressure gradients along the longitudinal direction generate longitudinal flow $u^\eta$ which pushes net baryon density outward. Hydrodynamic evolution with a larger net baryon diffusion coefficient leads to transporting more net baryon density to the central space-time rapidity region. This is because the net baryon diffusion current $q^\mu$ is driven by the spatial gradient of net baryon chemical potential, $\nabla^\mu (\mu_B/T)$. The double hump structure of the net baryon density distribution drives $q^\mu$ to point opposite to the longitudinal hydrodynamic flow and diffuses net baryon density back to mid-rapidity. The right panel of Fig.~\ref{fig1} shows the effects are seen in the final net proton rapidity distribution. With a larger baryon diffusion constant, more net protons (or net baryons) are transported from the forward and backward rapidities to the central rapidity region. Hence, the experimentally measured net proton rapidity distribution can in principle be used to constrain the size of the net baryon diffusion constant in the phenomenological study. However, such a phenomenological extraction is entangled with the initial shape of the net baryon density distribution.

At the RHIC BES energies, the time for the two nuclei to pass through each other is $1-3$ fm/c. Because there is no well-established effective theory to describe this stage of the evolution, we expect a substantial theory uncertainty on the system's dynamical evolution during this pre-equilibrium period. Thus, in order to build more realistic initial conditions for heavy-ion collisions at the RHIC BES energies and explore the effects of this pre-equilibrium stage on hadronic observables, we develop a generalized 3D Glauber-Lexus model \cite{Shen:2017yyy}. An earlier version of this model was introduced in Ref. \cite{Monnai:2015sca,Denicol:2015nhu} to study the effect of longitudinal fluctuation at the top RHIC and LHC energies. In this work, we extend the model to take into account the finite thickness of the incoming colliding nuclei and track the collision time and position for every binary collision. Strings that carry energy are produced from these binary collisions and are evolved assuming free-streaming for $\tau_\mathrm{th} = 0.5$ fm/c before they are considered to be thermalized. The detailed implementation of the model can be found in Ref. \cite{Shen:2017yyy}.
 %
\begin{figure}[ht!]
  \centering
  \begin{tabular}{cc}
  \includegraphics[width=0.4\linewidth]{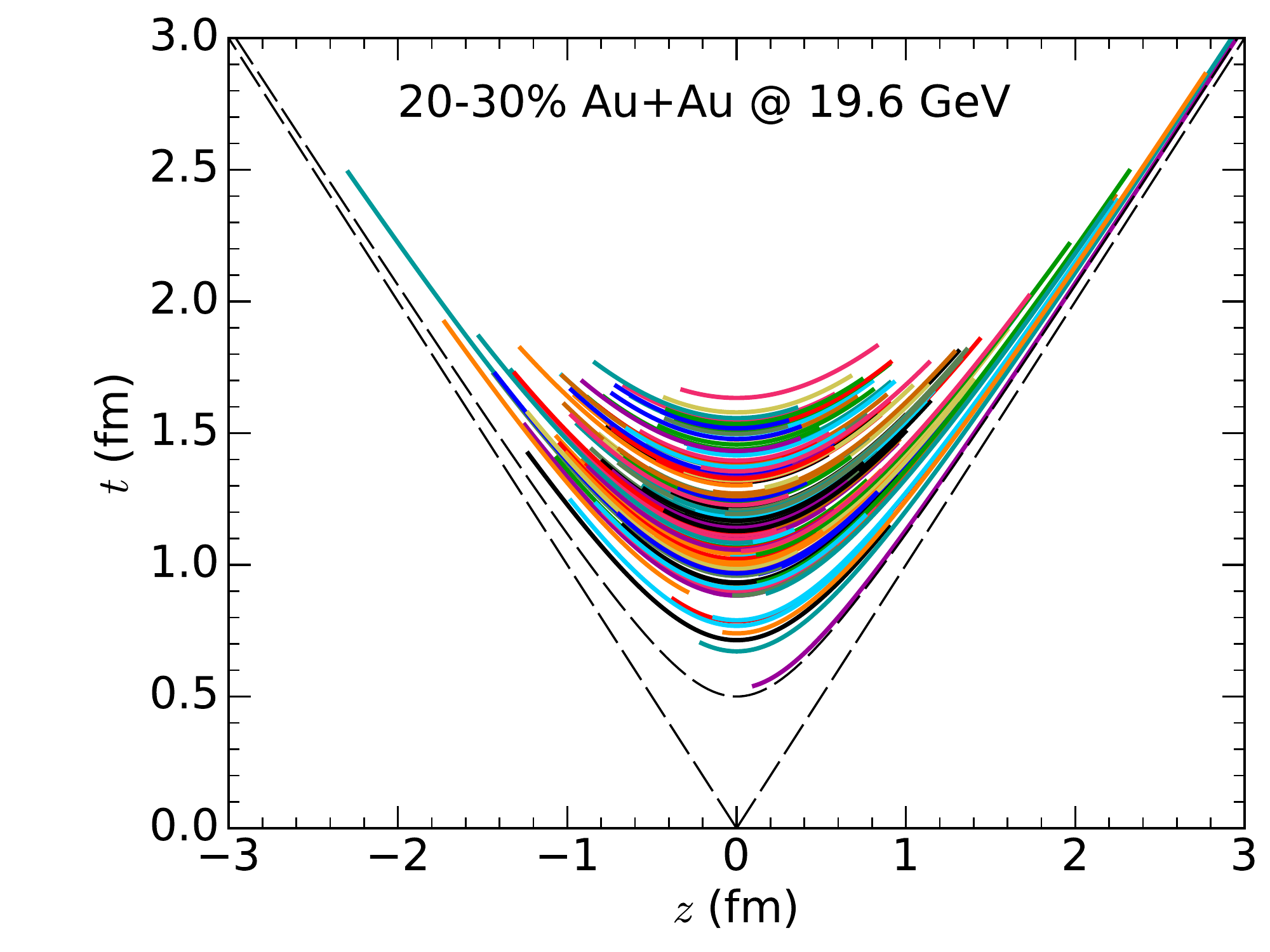} &
  \includegraphics[width=0.4\linewidth]{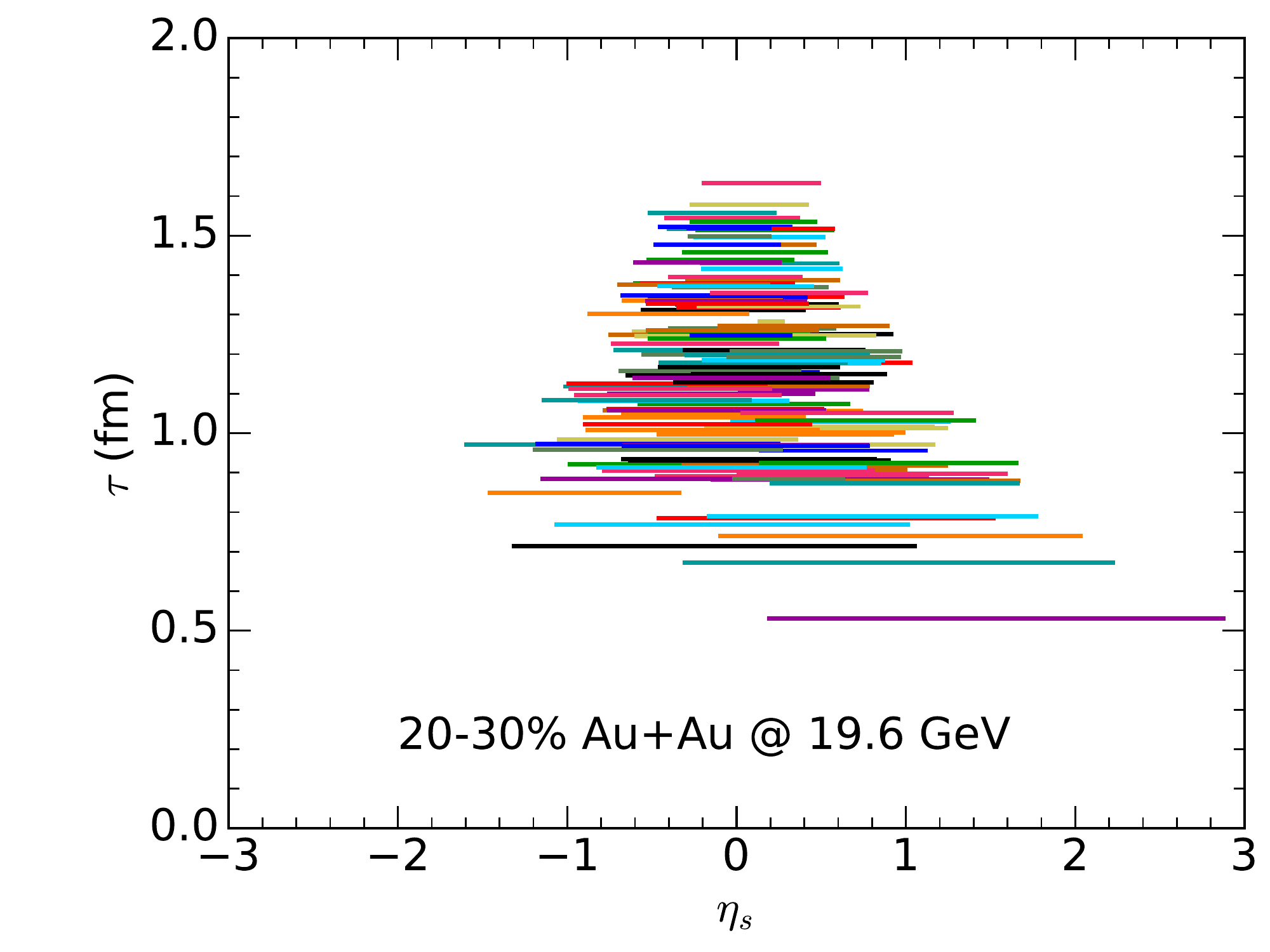}
  \end{tabular}
   \caption{The space-time picture of thermalized strings in one 20-30\% centrality Au+Au collision event at 19.6 GeV. Left panel is in $t-z$ coordinates and right panel is in $\tau-\eta_s$ coordinates. }
  \label{fig2}
\end{figure}
%
Fig.~\ref{fig2} shows the space-time distribution of where the strings become thermalized with hydrodynamic medium in one Au+Au collision in the 20-30\% centrality bin at 19.6 GeV. Please note that the thermalization times of the strings have a finite spread from $\tau = 0.5$ to 1.5 fm/c. This means that some early produced strings are thermalized with the hydrodynamic medium while other strings are just or not yet produced because of the finite thickness of the colliding nuclei. We checked that the spread of the thermalization time is negligible at 200 GeV and above. The finite spread of the thermalization times at low collision energies requires us to treat them as individual sources to dynamically initialize the hydrodynamic energy momentum tensor and net baryon current in our simulation
\begin{equation}
\partial_\mu T^{\mu\nu} = J_\mathrm{source}^\mu, \qquad \partial_\mu J^{\mu} = \rho_\mathrm{source}.
\end{equation} 
A similar approach was recently proposed in Ref. \cite{Okai:2017ofp}.

\section{Conclusion}
\label{conclusion}

In this work we study the effect of net baryon diffusion on hadronic observables. Hydrodynamic evolution with a larger net baryon diffusion constant leads to more net baryon number transport from the forward (backward) rapidity to mid-rapidity. The net proton rapidity distribution can be used to constrain the net baryon diffusion constant. However, such an extraction is entangled with the initial distribution of the net baryon density. To build a more realistic initial condition, we propose a dynamical initialization scheme to bridge the hydrodynamic evolution and pre-equilibrium stage. 

\bigskip
\noindent {\bf Acknowledgements}
This work was supported in part by the Natural Sciences and Engineering Research Council of Canada and in part by the U.S. Department of Energy, Office of Science, Office of Nuclear Physics, under Contract No. DE-SC0012704, and within the framework of the Beam Energy Scan Theory (BEST) Topical Collaboration.
BPS acknowledges a DOE Office of Science Early Career Award. AM is supported by JSPS Overseas Research Fellowships.
Computations were made in part on the supercomputer Guillimin from McGill University, managed by Calcul Qu\'ebec and Compute Canada. The operation of this supercomputer is funded by the Canada Foundation for Innovation (CFI), NanoQu\'ebec, RMGA and the Fonds de recherche du Qu\'ebec - Nature et technologies (FRQ-NT).





\bibliographystyle{elsarticle-num}
\bibliography{ref}






\end{document}